# A Sinking Approach to Explore Arbitrary Areas in Free Energy Landscapes

Zhijun Pan[1], Maodong Li[1], Dechin Chen[1], and Yi Isaac Yang[1,*]

1. Institute of Systems and Physical Biology, Shenzhen Bay Laboratory, Shenzhen 518132, China

* To whom correspondence should be addressed: yangyi@szbl.ac.cn

## Abstract

To address the time-scale limitations in molecular dynamics (MD) simulations, numerous enhanced sampling methods have been developed to expedite the exploration of complex free energy landscapes. A commonly employed approach accelerates the sampling of degrees of freedom associated with pre-defined collective variables (CVs), which typically tends to traverse the entire CV range. However, in many scenarios, the focus of interest is on specific regions within the CV space. This paper introduces a novel "sinking" approach that enables enhanced sampling of arbitrary areas within the CV space. We begin by proposing a gridded convolutional approximation that productively replicates the effects of metadynamics, a powerful CV-based enhanced sampling technique. Building on this, we present the SinkMeta method, which "sinks" the interior bias potential to create restraining potential "cliffs" at the grid edges. This technique can confine the exploration of CVs in MD simulations to a preset area. Our experimental results demonstrate that SinkMeta requires minimal sampling steps to estimate the free energy landscape for CV subspaces of various shapes and dimensions, including irregular two-dimensional regions and one-dimensional pathways between metastable states. We believe that SinkMeta will pioneer a new paradigm for sampling partial phase spaces, especially offering an efficient and flexible solution for sampling minimum free energy paths in high-dimensional spaces.

# Introduction

With the advancement of computing power and the widespread adoption of computational techniques, molecular dynamics (MD) simulations have found extensive applications across various scientific disciplines. The essence of these atomistic and molecular simulations lies in sampling, which involves calculating the probability distributions of the physical properties or processes of interest within the simulation system. Specifically, computing the free energy surfaces (FES) corresponding to the collective variables (CVs) of interest is crucial for studying the system's thermodynamics. However, even with powerful supercomputers like Anton[1], the time scales currently achievable *in silico* often fall short of the requirements for computing complex FES.

As a result, enhanced sampling methods[2-3] to overcome these time scale limitations have become an essential component of MD simulations. Some enhanced sampling approaches facilitate the global acceleration of all degrees of freedom (DOFs) within the simulation system, such as replica-exchange molecular dynamics (REMD) [4-5], simulated tempering[6], and integrated tempering sampling (ITS) [7-8]. These methods do not require prior setup of any CV and are relatively easy to use but typically offer limited acceleration for the physical processes in a specific partial phase space. Another class of enhanced sampling techniques is based on pre-defined CVs, such as umbrella sampling[9], local elevation[10], metadynamics (MetaD)[11], and variationally enhanced sampling (VES)[12]. These methods accelerate only the DOFs associated with the CVs, generally resulting in higher sampling efficiency for specific physical properties.

MetaD[11] is a powerful and widely used enhanced sampling approach that explores the FES by introducing a history-dependent bias potential into the simulation system. The MetaD method has spawned several variants, with well-tempered metadynamics (WT-MetaD) [13] being the most significant improvement, addressing the convergence problem of the bias potential by adaptively adjusting the height of the Gaussian kernel accumulated at each step. MetaD can also be combined with CV-free enhanced sampling methods, such as multiple walkers metadynamics (MW-MetaD) [14] as well as bias exchange metadynamics (BE-MetaD) [15-16] in conjunction with REMD and MetaITS[17-18] in combination with ITS.

Despite the various new techniques, CV-based enhanced sampling methods like MetaD are still limited by the size of the CV space they can explore. The phase space that can be visited within a given number of simulation steps is finite, causing the required simulation time for sampling to grow almost exponentially with increasing CV dimensions. However, in many cases, it is unnecessary to search the entire CV space. When one needs to sample a specific region of the FES, additional restraining potentials must be introduced. Unfortunately, adding suitable restraining potentials to high-dimensional CV spaces to control the CV within a particular area is extremely challenging.

In this article, we propose a "sinking" approach adapted from MetaD that allows sampling the FES in arbitrary areas of the simulation system. We first introduce a gridded convolutional approximation that efficiently achieves the equivalent enhanced sampling effect as the MetaD method. Then, we construct a sinking bias potential based on this convolutional approach. This method automatically creates "cliffs" of restraining potentials at the edges of pre-defined CV grids, thus limiting the sampling of CVs to desired areas. Finally, we present some examples of this approach for sampling specific regions with different shapes and dimensions in CV space and calculating their FES.

## Methodology

### A. Convolutional Metadynamics

Metadynamics (MetaD)[11] is an enhanced sampling method based on collective variables (CVs). CVs $s(R)$ is a set of functions of the atomic coordinates $R$ of the system, which can describe the physical behaviour of interest [19]: $s(R) = \{s_1(R), s_2(R), ..., s_D(R)\}$. The MetaD method achieves enhanced sampling by continuously accumulating Gaussian-type repulsive potentials $\{G(s(R); t)\}$ in the space of CVs $s(R)$ into the bias potential $V(s; t)$:

$$V(s(R); t) = \sum_t G(s(R); t) \tag{1}$$

$$G(s(R); t) = \omega(t) e^{-\frac{1}{2}\left\|\frac{s(R) - s'(t)}{\sigma}\right\|^2} \tag{2}$$

where $s'(t)$ is the value of the CVs $s(R)$ at the simulation step $t$, and $\omega(t)$ as well as $\sigma$ is the weight coefficient and standard deviation of the Gaussian function, respectively. The original

MetaD method has a constant weight coefficient $w$, whereas the more popular well-tempered metadynamics (WT-MetaD)[13] uses a function $\omega(t)$ that decreases gradually according to the previously accumulated bias potential $V(\boldsymbol{s}; t-1)$:

$$\omega(t) = w e^{-\left(\frac{1}{\gamma-1}\right)\beta V(\boldsymbol{s}'(t); t-1)} \tag{3}$$

where $\beta = k_B T$, $k_B$ is the Boltzmann constant and $T$ is the simulation temperature, $\gamma > 1$ is a bias factor constant. When $\gamma \to +\infty$, $\omega(t) = w$ is a constant value, i.e., equivalent to the original MetaD.

If using equation (1) to update the bias potential $V(\boldsymbol{s}, t)$ in the program, the single-step computational consumption of MetaD will increase continuously with the growing number of Gaussian kernels $\{G(\boldsymbol{s}(\boldsymbol{R}); t)\}$. A common solution is to accumulate the sums of Gaussian kernels $G(\boldsymbol{s}(\boldsymbol{R}); t)$ on $N$ pre-defined CV-grids $\{\boldsymbol{s}_i\}$,[20] which is used by many software packages such as PLUMED[21] and COLVARS[22].

Here, we introduce a novel gridded convolutional approach. The convolution of two "small" Gaussian functions with standard deviation $\sigma'$ is equal to a "large" Gaussian function with standard deviation $\sigma = \sqrt{2}\sigma'$: $\int_{-\infty}^{+\infty} d\xi \exp\left\{-\frac{(\xi-\mu)^2}{2(\sigma')^2}\right\} \exp\left\{-\frac{(x-\xi)^2}{2(\sigma')^2}\right\} = \sigma'\sqrt{\pi} \exp\left\{-\frac{(x-\mu)^2}{2\sigma^2}\right\}$. Therefore, we can construct a set of Gaussian-form basis functions with standard deviation $\boldsymbol{\sigma}' = \boldsymbol{\sigma}/\sqrt{2}$ on the CV-grids $\{\boldsymbol{s}_i\}$ to fit the Gaussian function $G(\boldsymbol{s}(\boldsymbol{R}); t)$ with standard deviation $\boldsymbol{\sigma}$:

$$G(\boldsymbol{s}(\boldsymbol{R}); t) \approx \frac{\omega(t)}{C(t)} \sum_i^N \Delta S_i e^{-\frac{1}{2}\left\|\frac{\boldsymbol{s}_i - \boldsymbol{s}'(t)}{\boldsymbol{\sigma}'}\right\|^2} e^{-\frac{1}{2}\left\|\frac{\boldsymbol{s}(\boldsymbol{R}) - \boldsymbol{s}_i}{\boldsymbol{\sigma}'}\right\|^2} \tag{4}$$

$$= \frac{\omega(t)}{C(t)} \sum_i^N \Delta S_i f_i(t) g_i(\boldsymbol{s}(\boldsymbol{R}))$$

$$f_i(t) = e^{-\frac{1}{2}\left\|\frac{\boldsymbol{s}_i - \boldsymbol{s}'(t)}{\boldsymbol{\sigma}'}\right\|^2} \tag{5}$$

$$g_i(\boldsymbol{s}(\boldsymbol{R})) = e^{-\frac{1}{2}\left\|\frac{\boldsymbol{s}(\boldsymbol{R}) - \boldsymbol{s}_i}{\boldsymbol{\sigma}'}\right\|^2} \tag{6}$$

$$C(t) = \frac{1}{(\sqrt{2})^D} \sum_i \Delta S_i f_i(t) \approx \prod_d^D (\sigma_d' \sqrt{\pi}) \tag{7}$$

where $\Delta S_i = \prod_d^D \Delta s_{i,d}$ is the product of each dimension of the $i$-th grid spacing $\Delta \boldsymbol{s}_i =$

$\{\Delta s_{i,1}, \Delta s_{i,2}, \ldots, \Delta s_{i,D}\}$, $g_i(s(\boldsymbol{R}))$ is the $i$-th basis functions, $f_i(t)$ is the weight coefficient of the basis function $g_i(s)$, and $C(t)$ is the normalisation factor. To ensure an accurate fit, any grid spacing $\Delta s_i$ should be less than half the standard deviation $\boldsymbol{\sigma}$: $\Delta s_i \leq 0.5\boldsymbol{\sigma}$. Figure 1a&b. show the schematic representations of the basis-functions $\{g_i(s)\}$, the weight coefficients $\{f_i\}$ and the fitted Gaussian function $G(s)$, respectively. See Appendix-A for detailed information about the gridded approximation for Gaussian convolution.

Therefore, we can change equation (1) for calculating the bias potential $V(s(\boldsymbol{R}); t)$ as:

$$V(\boldsymbol{s}(\boldsymbol{R}); t) \approx \sum_t \frac{\omega(t)}{C(t)} \sum_i^N \Delta S_i f_i(t) g_i(\boldsymbol{s}(\boldsymbol{R})) \tag{8}$$

$$= \sum_i^N k_i(t) \Delta S_i e^{-\frac{1}{2}\left\|\frac{\boldsymbol{s}(\boldsymbol{R}) - \boldsymbol{s}_i}{\boldsymbol{\sigma}'}\right\|^2}$$

$$= \sum_i^N G_i(\boldsymbol{s}(\boldsymbol{R}); t)$$

$$k_i(t) = \sum_t \frac{\omega(t)}{C(t)} f_i(t) \tag{9}$$

where $G_i(\boldsymbol{s}(\boldsymbol{R}); t) = k_i(t) \Delta S_i \exp\{-\frac{1}{2}\|\frac{\boldsymbol{s}(\boldsymbol{R}) - \boldsymbol{s}_i}{\boldsymbol{\sigma}'}\|^2\}$ is the Gaussian kernel at the $i$-th CV-grid, and $k_i(t)$ is the weight parameter of kernel $G_i(\boldsymbol{s}(\boldsymbol{R}); t)$. In this way, the updating of the bias potential $V(s, t)$ is performed by varying a fixed number of weight parameters $\{k_i(t)\}$, so the number of Gaussian kernels $\{G_i(\boldsymbol{s}(\boldsymbol{R}); t)\}$ need to be calculated at each step $t$ does not increase over the simulation time.

In practice, since the Gaussian is a local basis function, it is sufficient to fit the bias potential $V(s)$ using only a few numbers of Gaussian kernels at the grids $\{s_i\}$ within a cutoff distance $s_\text{cut}$ from the CV $s$, rather than using all the kernels $\{G_i(\boldsymbol{s}(\boldsymbol{R}); t)\}$:

$$V(\boldsymbol{s}(\boldsymbol{R}); t) \approx \sum_i^{|(\boldsymbol{s}(\boldsymbol{R}) - \boldsymbol{s}_i)_d| \leq s_{\text{cut},d}} k_i(t) \Delta S_i e^{-\frac{1}{2}\left\|\frac{\boldsymbol{s}(\boldsymbol{R}) - \boldsymbol{s}_i}{\boldsymbol{\sigma}'}\right\|^2} \tag{10}$$

where $|(\boldsymbol{s}(\boldsymbol{R}) - \boldsymbol{s}_i)_d| \leq s_{\text{cut},d}$ means that the absolute value of the component of $\boldsymbol{s}(\boldsymbol{R}) - \boldsymbol{s}_i$ in each dimension $d$ is less than or equal to $s_\text{cut}$. And when updating the weight parameters $k_i(t)$ using equation (9), it is also acceptable to accumulate the weight coefficients $f_i(t)$ only on the CV-grids $\{s_i\}$ that are within the cutoff distance $s_{\text{cut}}$.

$$f_i(t) \approx \begin{cases} e^{-\frac{1}{2}\left\|\frac{s_i - s'(t)}{\sigma'}\right\|^2}, & \left|(s_i - s'(t))_d\right| \leq s_{\text{cut},d} \\ 0, & \text{otherwise} \end{cases} \quad (11)$$

The cutoff distance $s_{\text{cut}}$ can be generally taken as $s_{\text{cut}} \geq 2.5\sigma$.

This convolutional metadynamics (ConvMeta) approach can fully reproduce the effects of enhanced sampling by the original or well-tempered MetaD, while also substantially decreases computational cost.

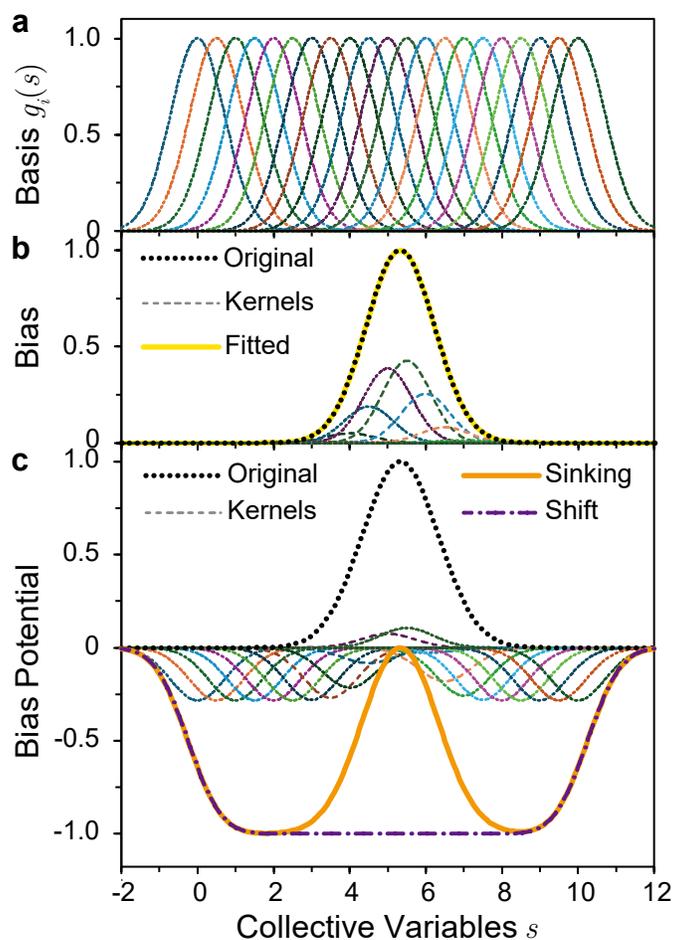

Figure 1. Schematic diagram of the bias potential of the ConvMeta and SinkMeta, where the sink depth $E_{\text{depth}}$ of the bias maximum is 0. a) The Gaussian-type basis-functions $\{g_i(s)\}$. b) Comparison of the bias potential (yellow solid line) fitted using a series of small Gaussian kernels $\{G_i(s)\}$ (colored dashed line) with the original Gaussian-type repulsive potential (black dotted line). c) The bias potential of SinkMeta. The blue dotted line is the bias potential $V(s(\boldsymbol{R});t)$ of the original MetaD, the orange solid line is the bias potential $V_{\text{sink}}(s(R);t)$ of SinkMeta, the coloured dashed line is the shifted Gaussian kernels $\{G'_i(s)\}$, and the violet dotted dashed line represents the shift potential $V_{\text{shift}}(s(R);t)$ i.e. the difference between the sinking and original bias potentials.

### B. "Sinking" Metadynamics

The bias potential $V(s(R))$ of MetaD that consists of the kernels of repulsive Gaussian $G(s(R))$ will drive the CV $s(R)$ away from the position it has already visited, thus forcing the sampling of the system to traverse the entire CV space. The size of the space to be explored increases exponentially with the dimension $D$ of the CV $s(R)$. Therefore, for adequate sampling with MetaD, the dimensionality of the CV $s(R)$ used must be very small, most commonly one or two, with three or above being relatively rare. Moreover, for non-periodic CV, the explorable space is theoretically infinite. Furthermore, it is also troublesome for the gridded approximation: once the CV $s(R)$ is squeezed out of the range of preset grids $\{s_i\}$, the approach will fail. Thus, using non-periodic CVs in gridded MetaD usually requires building additional "walls" at the grid boundaries, i.e., restraining potentials for preventing CV from crossing the limit. However, the restraining potential of high-dimensional CVs is complicated to design.

Here, we propose a "sinking" approach adapted from the previously introduced ConvMeta approach, which uses the kernels of attractive potential to estimate the bias potential $V(s(R))$. If a shift factor $v_{\text{shift}}(t)$ is subtracted for each of the weighting parameters $\{k_i(t)\}$ in equation (8), then we will obtain a "sinking" bias potential $V_{\text{sink}}(s(R);t)$:

$$V_{\text{sink}}(s(R);t) = \sum_i^N [k_i(t) - v_{\text{shift}}(t)]\Delta S_i e^{-\frac{1}{2}\left\|\frac{s(R)-s_i}{\sigma'}\right\|^2} \quad (12)$$
$$= \sum_i^N G'_i(s(R);t)$$

where $G'_i(s(R);t) = [k_i(t) - v_{\text{shift}}(t)]\Delta S_i \exp\{-\frac{1}{2}\|\frac{s(R)-s_i}{\sigma'}\|^2\}$ is the shifted Gaussian kernel at the $i$-th CV-grid. We hope the maximum value of the sinking bias potential $V_{\text{sink}}(s(R);t)$ at any time $t$ to be equal to a constant value $-E_{\text{depth}}$ that is less than 0, where $E_{\text{depth}}$ is the sinking depth of the bias potential maximum $V_{\max}(t)$. Therefore, the shift factor $v_{\text{shift}}(t)$ is:

$$v_{\text{shift}}(t) = \frac{V_{\max}(t) + E_{\text{depth}}}{\prod_d^D (\sigma'_d \sqrt{2\pi})} = \frac{V_{\text{depth}}(t)}{C'} \tag{13}$$

$$V_{\text{depth}}(t) = V_{\max}(t) + E_{\text{depth}} \tag{14}$$

where $C' = \prod_d^D (\sigma'_d \sqrt{2\pi})$ is the normalisation constant, $V_{\max}(t)$ is the maximum value of the original bias potential $V(s(\boldsymbol{R}); t)$, and $V_{\text{depth}}(t)$ is the depth to which the interior region of the bias potential $V(s(\boldsymbol{R}), t)$. See Appendix-B for detailed information about the shift factor.

We call this method SinkMeta. Compared to the original MetaD, it is equivalent to introducing an additional shift potential $V_{\text{shift}}(s(\boldsymbol{R}); t)$ to the bias potential:

$$\begin{aligned}
V_{\text{shift}}(s(\boldsymbol{R}); t) &= V_{\text{sink}}(s(\boldsymbol{R}); t) - V(s(\boldsymbol{R}); t) \\
&= -v_{\text{shift}}(t) \sum_i^N \Delta S_i e^{-\frac{1}{2} \left\| \frac{s(\boldsymbol{R}) - s_i}{\sigma'} \right\|^2} \\
&= -v_{\text{shift}}(t) \Phi(s(\boldsymbol{R}))
\end{aligned} \tag{15}$$

where $\Phi(s(\boldsymbol{R})) = \sum_i^N \Delta S_i \exp\left(-\frac{1}{2} \left\| \frac{s(\boldsymbol{R}) - s_i}{\sigma'} \right\|^2\right)$ the cumulative function of Gaussian basis $\{g_i(s(\boldsymbol{R}))\}$. As shown in Figure 1c, most of the Gaussian kernels $G'_i(s(\boldsymbol{R}); t)$ become attractive potentials, and the bias potential $V_{\text{sink}}(s(\boldsymbol{R}))$ in the interior region $S_{\text{inter}}$ of the CV grids $\{s_i\}$ shifted by a constant value $V_{\max}$ compared to the original bias potential $V(s(\boldsymbol{R}))$. This means that the system will be affected by the same bias force at this region as the original bias force $\boldsymbol{F}(\boldsymbol{R}; t) = -\partial V(s(\boldsymbol{R}))/\partial \boldsymbol{R}$ before sinking. While at the margins of the CV grids $\{s_i\}$, the bias potential $V_{\text{sink}}(s(\boldsymbol{R}))$ form cliff-like restraining potentials, whose presence confines the sampling of the CV $s(\boldsymbol{R})$ in the interior of the grids $\{s_i\}$ and prevents it from crossing the boundary. See Appendix-C for detailed information about the boundary effect.

Therefore, SinkMeta can achieve the same sampling effect inside the CV grids as the ordinary WT-MetaD approach and, more importantly, without the risk of the CV escaping the grids after a long simulation time. It implies that we can set up irregular CV-grids in SinkMeta to achieve enhanced sampling of arbitrary regions of the free energy surface.

### C. Thermodynamics Calculation

The calculation of thermodynamic properties using SinkMeta is analogous to the WT-MetaD approach[13]. After the iterative convergence of the bias potential $V_{\text{sink}}(s)$, the free energy $F(s)$ corresponding to CV $s(R)$ at the interior region $S_{\text{inter}}$ of CV-grids $\{s_i\}$ can be calculated as:

$$F(s) \propto -\left(\frac{\gamma}{\gamma-1}\right) V_{\text{sink}}(s), \qquad (s \in S_{\text{inter}}) \tag{16}$$

Note that this formula cannot be used to compute the free energy of CVs outside the grids $\{s_i\}$, and calculating CVs inside the grids but at the margins will also cause errors. Instead, the Boltzmann distribution $p_0(R)$ of any observables in the system can be calculated as follows:

$$p_0(R) = p(R) e^{\beta\{V_{\text{sink}}[s(R);t] - c(t)\}} \tag{17}$$

$$\begin{aligned} c(t) &= \frac{1}{\beta} \log \frac{\int ds\, e^{-\beta F(s)}}{\int ds\, e^{-\beta[F(s)+V_{\text{sink}}(s;t)]}} \\ &\approx \frac{1}{\beta} \log \frac{\int ds\, e^{\frac{\gamma}{\gamma-1}\beta V(s;t)}}{\int ds\, e^{\frac{1}{\gamma-1}\beta V(s;t)} e^{-\beta V_{\text{shift}}(s;t)}} \end{aligned} \tag{18}$$

where $p(R)$ is the sampling probability obtained from MD simulations, $c(t)$ is the revised factor[23-24] for weights.

### D. Code Available

We have implemented the ConvMeta and SinkMeta methods in the MD simulation software SPONGE[25] and MindSPONGE[26]. The relevant code can be downloaded from the Gitee Code Repository: https://gitee.com/d2denis/cudasponge-pan (SPONGE) and https://gitee.com/helloyesterday/mindsponge/tree/develop/ (MindSPONGE). The technical details and benchmark tests against well-established software can be found in the Supplementary Material.

## Results

**Sampling of the Complete CV space**

First, we tested the ConvMeta and SinkMeta methods for sampling the entire CV space using a typical alanine dipeptide (ACE-ALA-NME) system in vacuum. This system was modelled using the AMBER FF19SB[27] force field. We selected the Ramachandran dihedral angles $\phi$ and $\psi$ as the CVs, which are divided into 50*50 uniform grids.

We performed the MD simulations using a modified version of SPONGE. Additionally, we conducted MD simulations with conventional WT-MetaD using the same hyperparameters, utilising the well-established software SANDER in AmberTools24[28] with the PLUMED2.7[21] plug-in library.

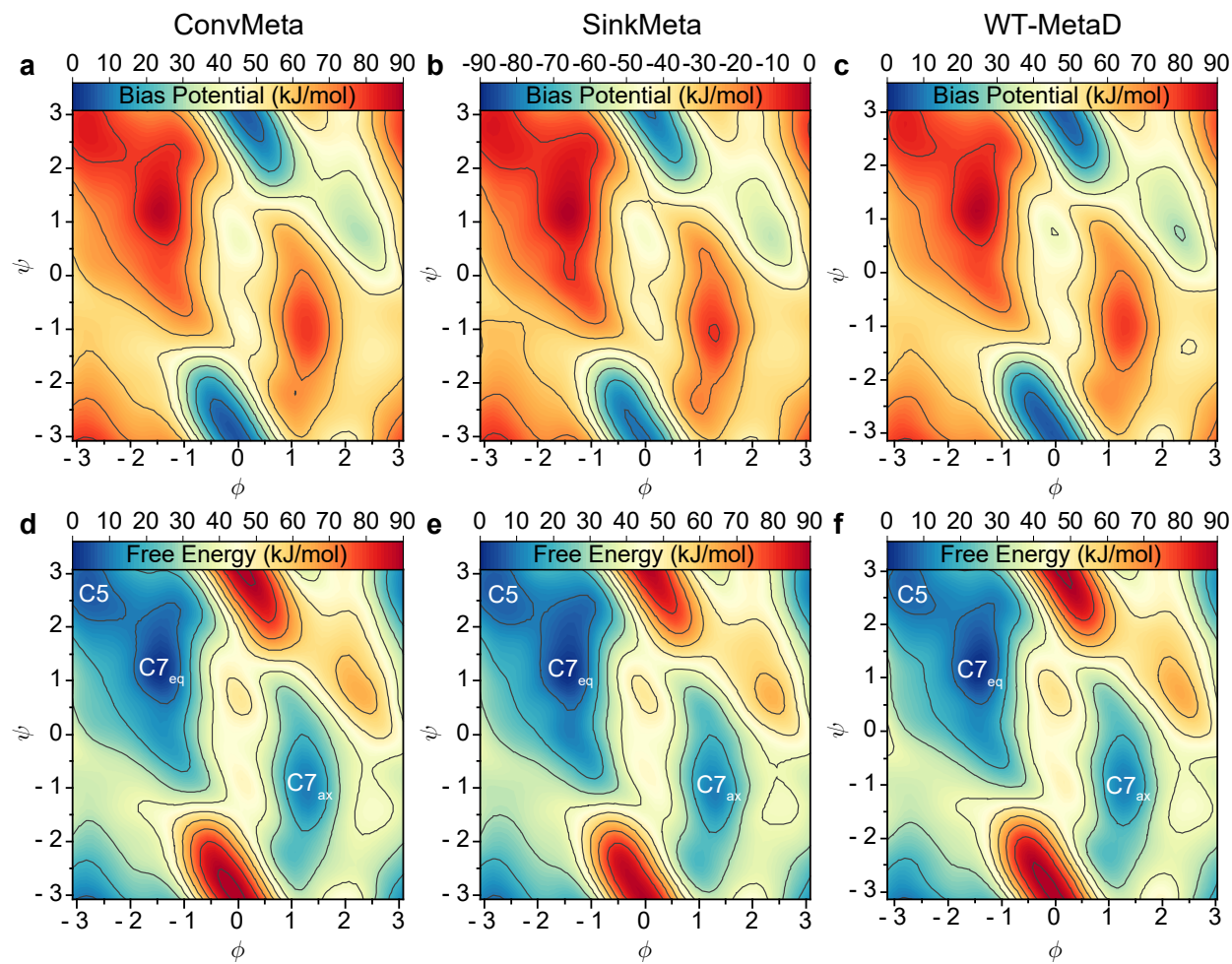

Figure 2. Landscape of the bias potential and free energy expressed as the function of the CVs $(\phi, \psi)$. a-c) The bias potential was obtained from ConvMeta, SinkMeta, and WT-MetaD, respectively. d-f) The free energy surface was calculated using conventional ConvMeta, SinkMeta, and WT-MetaD, respectively.

Figure 1 shows the landscape of bias potentials obtained using ConvMeta, SinkMeta, and WT-

MetaD. The results indicate that ConvMeta produces a bias potential identical to that of WT-MetaD, demonstrating its consistency with the conventional MetaD approach. The bias potential landscape generated by SinkMeta is also consistent with that of traditional MetaD, except that its values are all negative. The free energy landscapes calculated using the three enhanced sampling methods are in perfect agreement (Figure 1d-f), all demonstrating the three metastable states C5, C7$_{eq}$, and C7$_{ax}$ of the alanine dipeptide in vacuum. This indicates that SinkMeta can achieve the same sampling effect as the classical MetaD method for periodic CVs without boundaries.

**Sampling for Specified CV Range**

Then, we verify the sampling effect of SinkMeta for specific (CV) ranges using a capped decamer of alanine (referred to as deca-alanine) in vacuum. Deca-alanine is known for its propensity to form α-helices and is commonly used in theoretical investigations of conformational equilibria of short peptide segments.[29] Extension of deca-alanine can lead to its reversible unfolding, so its end-to-end distance $r_{ee}$ is often used as the CV for enhanced sampling. Deca-alanine can transform into various states at different end-to-end distances $r_{ee}$ (Figure 3a), which covers a large variable range. Traditionally, upper and lower "walls" (restraining potentials) must be added to the bounds to sample a specific range of a CV of distance.

Instead, with SinkMeta, all that is needed is a preset sampling range, allowing us to sample the distance $r_{ee}$ within that range in the MD simulation without any additional operations. Figure 3b shows the sampling effects for $r_{ee}$ with different CV value ranges set in SinkMeta. It is evident that the sampling of $r_{ee}$ in the MD simulation is strictly limited to a predefined range of values without crossing the boundary. Figure 3c presents the results of free energy calculations for different preset ranges of CV, implying that SinkMeta can estimate the free energy for a local phase space like umbrella sampling but with more convenience and flexibility in selecting the sampling space.

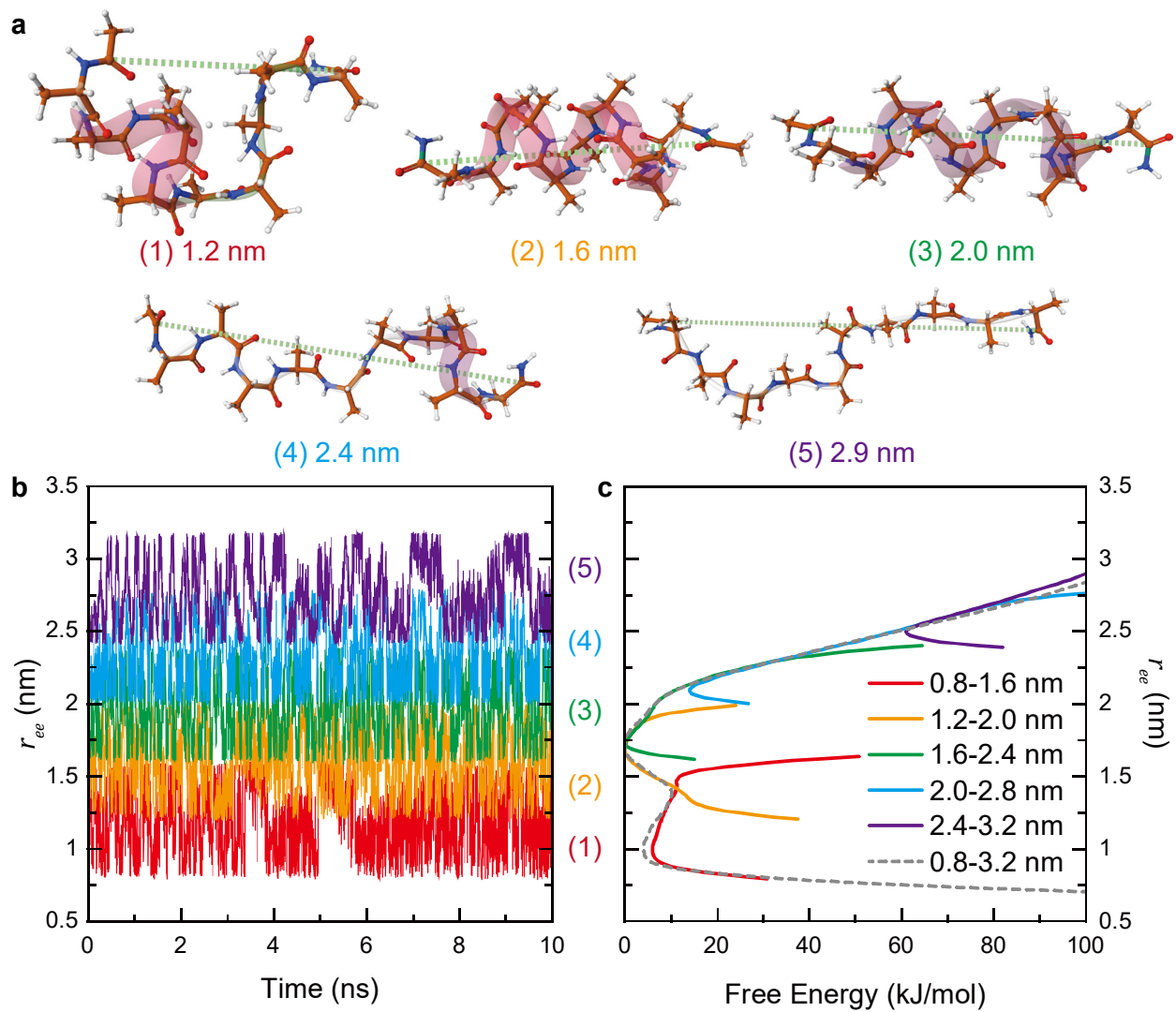

Figure 3. a) Conformations of Deca-alanine at different end-to-end distances. b) Evolution of end-to-end distances in MD simulations with different CV ranges of SinkMeta. c) Free energy surface as a function of end-to-end distance calculated using different CV ranges of SinkMeta.

# Sampling Irregular Area in CV Space

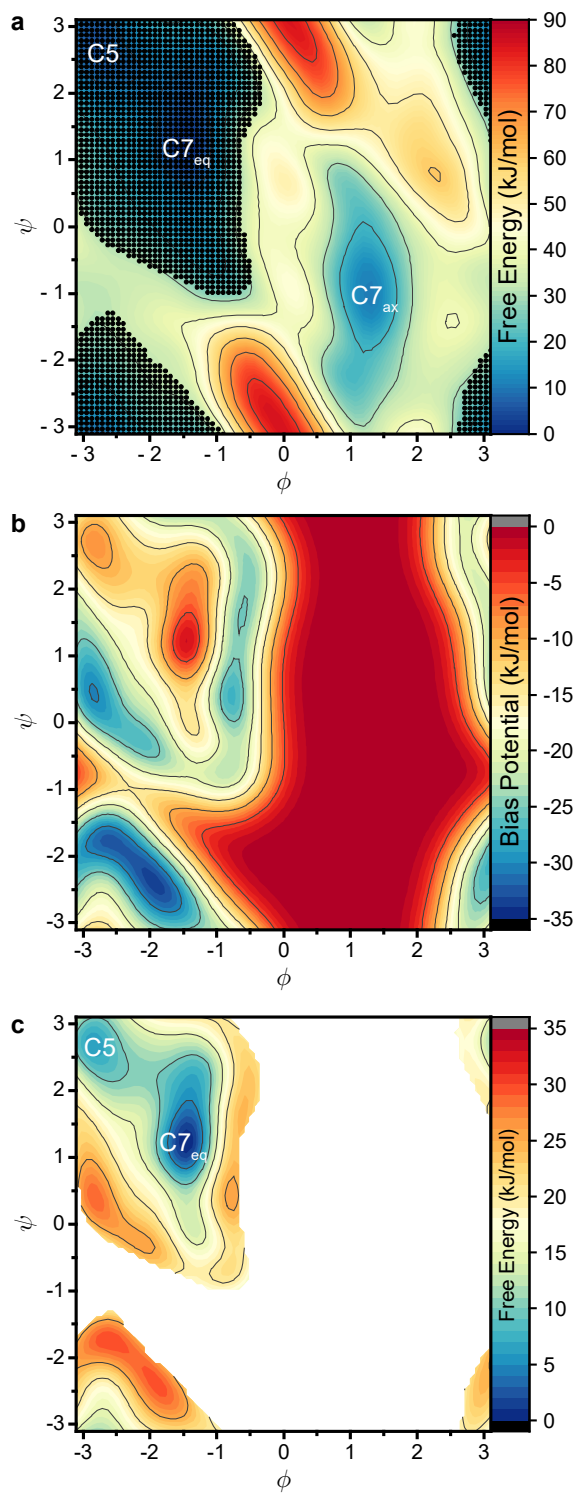

Figure 4.  Enhanced sampling for regions where C5 and C7$_{eq}$ of alanine dipeptide are located. a) The position of Gaussian basis functions in the CV space. b) The bias potential generated by SinkMeta. c) Free energy surface as the function of $(\phi, \psi)$ calculated using SinkMeta.

Next, we examine the sampling effectiveness of SinkMeta for arbitrary regions of CV space. We used a typical alanine dipeptide system *in vacuo* and utilised the dihedral angles $(\phi, \psi)$ as CVs. This system has three metastable states: C5, C7$_{eq}$, and C7$_{ax}$, where C5 and C7$_{eq}$ are neighbouring states in phase space (Figure 4a). Using conventional MetaD with these CVs would sample the entire $(\phi, \psi)$ space, resulting in a free energy landscape like the one in Figure 2d-f. In contrast, the SinkMeta method allows us to sample only a specific region of that space.

Here, we used SinkMeta to sample an irregular region where the alanine dipeptide's metastable states C5 and C7$_{eq}$ are located. We put the Gaussian basis functions only in areas around C5 and C7$_{eq}$, as illustrated in Figure 4a, and confined the sampling to these areas in the MD simulation. The resulting bias potential (Figure 4b) shows a "sinking" landscape, with zero potential outside the sampling region and negative value within it. This sinking bias potential restrains the sampling to predefined Gaussian grids, thus preventing it from escaping to other areas during the MD simulation. Figure 4c presents the free energy surface calculated using SinkMeta, demonstrating the successful reconstruction of the local free energy landscape around the C5 and C7$_{eq}$ states.

**Sampling Path in CV Space**

The SinkMeta approach is highly flexible in choosing sampling areas, which can be subspace blocks or even reduced to one-dimensional paths in CV spaces. Again, we used the alanine dipeptide system to demonstrate the effects of SinkMeta on path sampling. Many path-searching methods, such as nudged elastic band (NEB) [30-32], string methods[33-36], and traveling-salesman-based automated path searching (TAPS) [37], have shown their ability to find the minimum free energy paths (MFEPs) of alanine dipeptide. Typically, these searched paths can serve as reaction coordinates for enhanced sampling methods like umbrella sampling to estimate the free energies on them. However, in many cases, it is not easy to define the paths as a few reaction coordinates that can be used as restraints. [31] In contrast, using SinkMeta requires only continuous and smooth points on the path in CV space.

We used SinkMeta to sample a path between the C7$_{eq}$ and C7$_{ax}$ states of the alanine dipeptide, where a high energy barrier separates these two dominant metastable states. We searched for an MFEP from C7$_{eq}$ to C7$_{ax}$ in the $(\phi, \psi)$ space using NEB and placed a series of Gaussian basis functions along this path (Figure 5a). In addition, we extended both ends to high free energy locations to avoid errors at the path margins, arranging 92 basis functions.

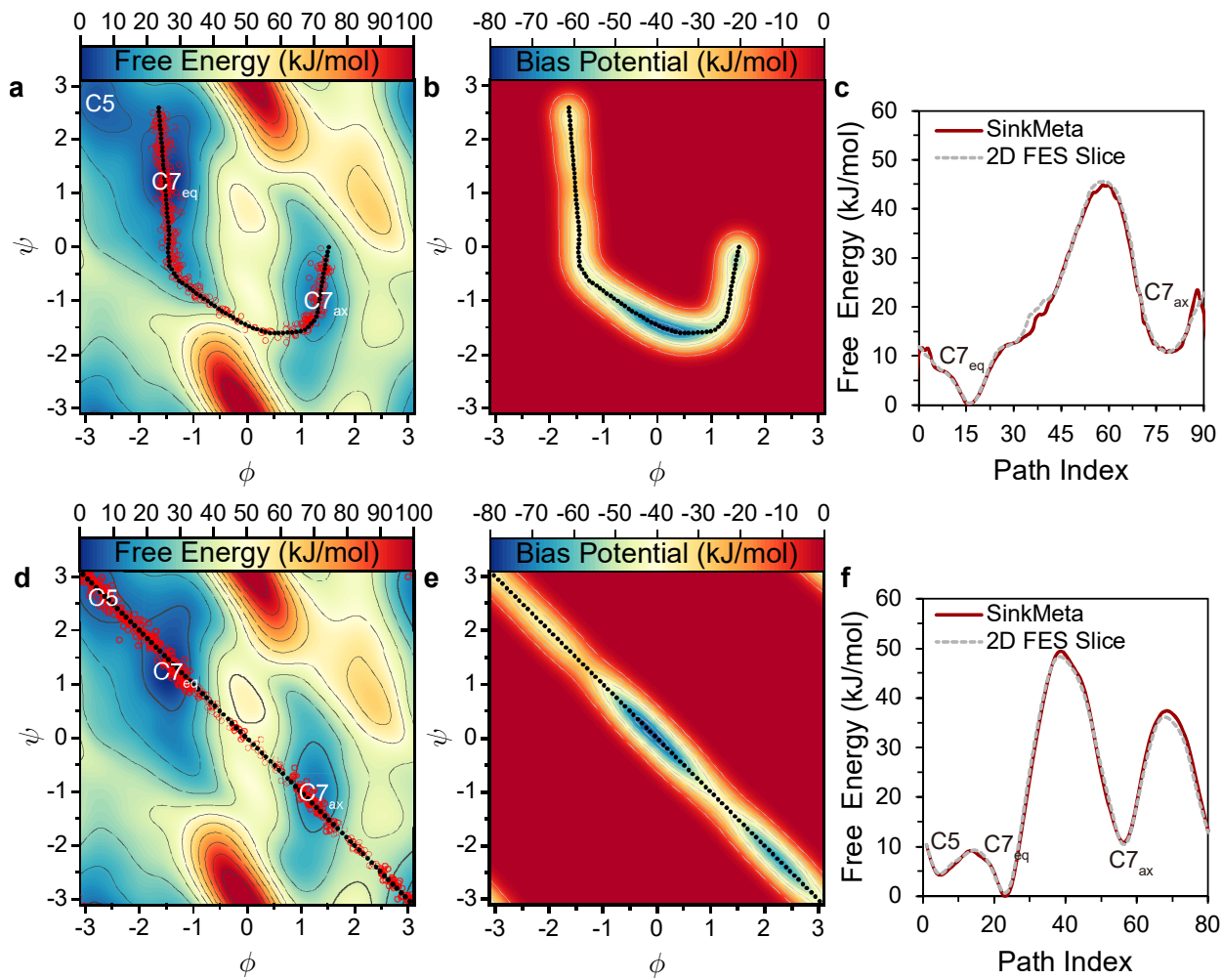

Figure 5. Sampling of paths in the free energy surface of alanine dipeptide using SinkMeta. Upper panel: sampling along the MFEP between C7$_{eq}$ and C7$_{ax}$ states. Lower panel: sampling along the diagonal path in the $(\phi, \psi)$ space. a&d) Sampling paths and footprints. The black dots are the location of Gaussian basis functions, and the red circles are the footprint left by the evolution of CV during MD simulation. b&e) Bias potentials in the $(\phi, \psi)$ space. c&e) Free energy surfaces with paths as reaction coordinates. The red solid line is the free energy calculated using the SinkMeta method with the path as the sampling region, while the grey dashed line is the sliced value of the 2D free energy surface over the path.

SinkMeta achieves highly efficient path sampling. As shown in Figure 5a, the evolution of the CVs remained concentrated near the path during the MD simulation, as the sinking bias potential attracted the CVs around the path (Figure 5b). This allows rapid transitions between $C7_{eq}$ and $C7_{ax}$ states in a short simulation time. After 1 ns of MD simulation, we calculated the system's FES as a function of the path (Figure 5c). The free energy calculated with SinkMeta is identical to the result from slicing the two-dimensional free energy surface obtained with conventional MetaD in a 50 ns simulation.

In fact, SinkMeta can sample any path on the free energy surface, not just the MFEP. We constructed Gaussian grids along the diagonal of the 2D FES, passing through C5, $C7_{eq}$, and $C7_{ax}$ states (Figure 5d). Figure 5e shows that SinkMeta forms a sinking bias potential along the diagonal, confining the CV variations of the system. We computed the free energy for the diagonal path, requiring a slightly longer sampling time due to higher energy barriers. The FES from a 5 ns simulation is shown in Figure 4f, demonstrating that SinkMeta can reproduce the values of diagonal slices of the 2D free energy surface. This proves that using SinkMeta can compute the free energy difference between different states on arbitrary paths, enabling transitions between the states instead of deliberately searching for MFEPs.

**Path Sampling in Cartesian Space**

Path sampling with SinkMeta can be applied to any CV space. Next, we show its effect in Cartesian space using a DNA-coumarin binding system. DNA is a pharmacological target for many drugs, and its binding to small molecules has multiple modes[38]. Compounds binding to the DNA minor groove have potential clinical utility against diseases like cancer and sleeping sickness.[39] A natural product, coumarin, can bind to the DNA minor groove with various pharmacological properties. Here, we investigated the binding mode of coumarin with a DNA duplex (sequence d(5'-GCGCATGCTACGCG-3')$_2$). We first used the blind docking software DSDP[40] to predict the binding site of coumarin to DNA, and the computation results shows that coumarin most prefers binding at the site between 9T and 10A in the DNA minor groove (see Figure 6a and PDB S1). This suggests that coumarin binding at the minor groove of DNA may be favoured for specific sequence locations.

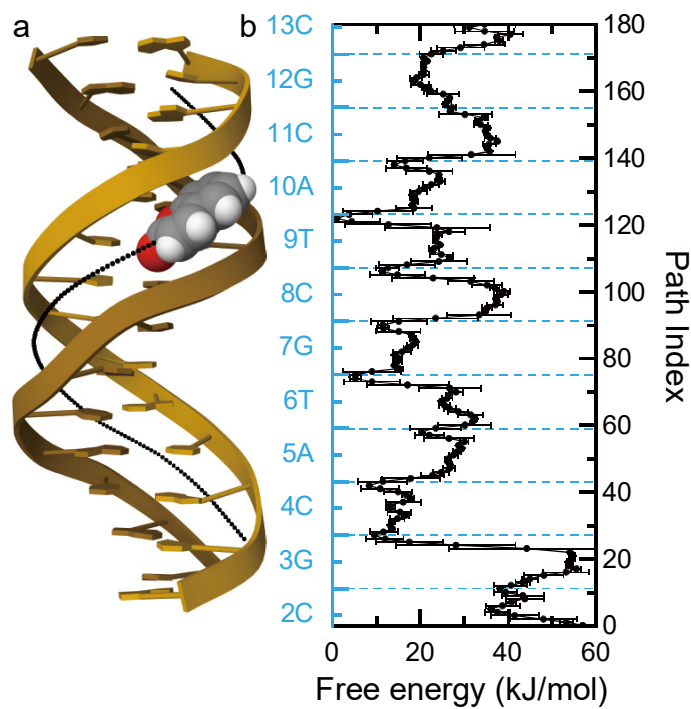

Figure 6. System of coumarin binding to DNA. a) 3D structure of coumarin molecules docked in the DNA minor groove. The black dots represent the positions of Gaussian basis functions for SinkMeta. b) Relative binding free energy along the path. The blue axis indicates the location of the path corresponding to the DNA sequence

We used SinkMeta to calculate the relative binding free energies of coumarin at different sites along the DNA minor groove. The Cartesian coordinate of the coumarin molecule's center of mass (COM) was selected as the CV for enhanced sampling. We set a uniform distribution of 181 Gaussian basis functions along the DNA minor groove's helix (Figure 6a), passing through the COM of the docked coumarin. Performing MD simulations with SinkMeta allowed coumarin to move back and forth along the minor grooves (see Movie S1). After 50 ns of simulation, we calculated the free energy landscape along the path, representing the relative binding free energies of coumarin at different DNA minor groove locations (Figure 6b).

The minimum relative binding free energy occurs at the site between bases 9T and 10A, consistent with docking results. Moreover, 9 of the 10 conformations with the highest binding affinity scores predicted by DSDP are located between 9T and 10A (see PDB S1), while one is between 6T and 7G, corresponding to the next lowest free energy position in Figure 6b. This implies that SinkMeta accurately reveals the relative binding free energies at different DNA minor groove sites. MD simulation with SinkMeta can also show the molecular details of drug-DNA binding, aiding the investigation of the physiological mechanism of coumarin-DNA interactions. Further studies on drug-DNA binding mechanisms will be presented in subsequent articles, as this article focuses on the SinkMeta methodology.

## Discussion and Conclusion

This paper introduces a novel approach, SinkMeta, in MD simulation, allowing enhanced sampling of arbitrarily shaped regions in the CV space. We propose the ConvMeta approximation to achieve an equivalent enhanced sampling effect as MetaD, but more efficiently. Based on this, we introduce a "sinking" bias potential with restraining "cliffs" at the edges, which effectively limit the exploration of the CV to the desired area, thus significantly reducing the simulation time required for sampling. We validated the sampling effectiveness of SinkMeta in systems of alanine dipeptide, deca-alanine, and coumarin binding to DNA. Our results indicate that SinkMeta can flexibly achieve efficient enhanced sampling of arbitrary CV areas, including specific CV ranges, irregular CV regions, and one-dimensional paths in high-dimensional CV spaces.

Using the ConvMeta and SinkMeta methods requires attention to grid edge effects. The gridded convolutional approach can be viewed as a particular case of Gaussian-mixture-based methods[41-43]. The Gaussian kernel functions with the same hyperparameters are uniformly distributed on the grids, allowing us to use equation [4] to calculate the Gaussian kernel based on the infinite integral. However, this condition is unmet at grid margins. Additionally, SinkMeta produces "cliffs" of restraining potentials at grid edges, which also affects free energy calculation. Therefore, the grid should include some redundancies by expanding the sampling area by at least $2\sigma$. Moreover, one should avoid placing outermost grids with prominent shapes (e.g., the two terminals of a one-dimensional path in high-dimensional CV space) in the free energy basin to prevent excessive bias potential accumulation.

The starting structure of the system should be within the grids when performing MD simulations with SinkMeta. SinkMeta only generates restraining force at grid margins, and the bias force far from the grids is zero. Thus, the initial CV in the simulation should not be outside the grids. Additionally, if the CV escapes the grid area during the simulation, it is necessary to increase the sink depth $E_{\text{depth}}$ to prevent this problem. We are also developing a new restraining potential to address this issue.

We believe this "sinking" approach can open a new paradigm for enhanced sampling. Traditionally, CV-based methods typically sample the entire CV space, which is unnecessary for many scenarios. SinkMeta pioneers a new way of sampling arbitrary areas of CV space. As a special case, SinkMeta is highly suitable for path sampling. Previous works have typically used umbrella sampling[44-46] or Path-CV[47-48] for path sampling. Compared to umbrella sampling[9], SinkMeta allows a continuous transition of CVs along the entire path. While Path-CV[49-50] transforms the points on the path into two CVs (i.e., the progress $s$ along the path and the distance $z$ from the path) and performs enhanced sampling on these 2D CVs $(s, z)$. In contrast, SinkMeta directly explores the CVs along the one-dimensional path in the original CV space. This means SinkMeta requires only very short simulation times to estimate the free energy landscape along the path accurately, making it an efficient and flexible method for path sampling.

The SinkMeta method also has the potential to be combined with path-searching methods and artificial intelligence (AI). The flexible grid settings in SinkMeta make it possible to achieve more efficient sampling by integrating path-searching methods. Recent advances in AI for molecular modeling and simulations [26, 51] have provided significant insights, especially deep reinforcement learning[52], which is well-suited for enhanced sampling[53]. If reinforcement learning can dynamically adjust SinkMeta grids during MD simulations, it is possible to realise intelligent simulation and sampling.

**Simulation Details**. The deca-alaine (ACE-ALA*10-NHE) system was modelled using AMBER FF14SB[54] force field. The height $w$ and standard deviation $\sigma$ of the Gaussian repulsive potential in MetaD are 1.67 kJ/mol and 0.04 nm, respectively. The well-tempered bias factor $\gamma$ for WT-MetaD is 50, and the sinking depth $E_{\text{depth}}$ for SinkMeta is 25 kJ/mol.

The alanine dipeptide (ACE-ALA-NME) system was modelled using AMBER FF19SB[27] force field. The height $w$ and standard deviation $\sigma$ of the Gaussian repulsive potential in MetaD are 2.0 kJ/mol and 0.314 rad, respectively. The well-tempered bias factor $\gamma$ for WT-MetaD is 10.

The sinking depth $E_{\text{depth}}$ of SinkMeta is 2.5 kJ/mol for the system of sampling 2D irregular regions and 25 KJ/mol for the system of sampling paths.

For the DNA-coumarin binding system, the DNA molecule was built using the PDB file with ID 2M2C[55] and modelled using the AMBER BSC1[56] force field, and the coumarin molecule was modelled using the AMBER GAFF[57] force field. The system was immersed in a periodic solvent box containing 10245 TIP3P[58] water molecules and 26 sodium ions, and the initial dimensions of the box is 6nm*6nm*8nm. The height $w$ and standard deviation $\sigma$ of the Gaussian repulsive potential in MetaD are 0.6 kJ and 0.05 nm, respectively. The well-tempered bias factor $\gamma$ for WT-MetaD is 20, and the sinking depth $E_{\text{depth}}$ for SinkMeta is 2.5 kJ/mol.

## Acknowledgement

The authors thank Yi Qin Gao, Xu Han, Yijie Xia, Haohao Fu and Lizhe Zhu for useful discussion. Computational resources were supported by Shenzhen Bay Laboratory supercomputing centre. This work was supported by the National Science and Technology Major Project (No. 2022ZD0115003) and the National Natural Science Foundation of China (22273061 to Y.I.Y).

# Appendix

## A. Gridded Approximation for Gaussian Convolution

We set the convolution of Gaussian functions $f(x) = g(x) = e^{-\frac{x^2}{2(\sigma')^2}}$ as $\mathcal{G}(x)$:

$$\mathcal{G}(x) = (f * g)(x) = \int_{-\infty}^{+\infty} d\tau f(\tau) g(x - \tau) \tag{S1}$$

Thus, the function $\mathcal{G}(x - \mu)$ is:

$$\mathcal{G}(x - \mu) = (f * g)(x - \mu) = \int_{-\infty}^{+\infty} d\tau f(\tau) g(x - \mu - \tau) \tag{S2}$$

Set $\xi = \tau + \mu$, then:

$$\begin{aligned}
\mathcal{G}(x - \mu) &= \int_{-\infty}^{+\infty} d\xi f(\xi - \mu) g(x - \xi) \\
&= \int_{-\infty}^{+\infty} d\xi e^{-\frac{(\xi - \mu)^2}{2(\sigma')^2}} e^{-\frac{(x - \xi)^2}{2(\sigma')^2}} \\
&= \int_{-\infty}^{+\infty} d\xi e^{-\frac{x^2 + \mu^2 - 2(x+\mu)\xi + 2\xi^2}{2(\sigma')^2}} \\
&= \int_{-\infty}^{+\infty} d\xi e^{\frac{(x-\mu)^2}{2(\sigma')^2}} e^{-\frac{\frac{(x+\mu)^2}{2} - 2(x+\mu)\xi + 2\xi^2}{2(\sigma')^2}} \\
&= e^{-\frac{(x-\mu)^2}{2(\sqrt{2}\sigma')^2}} \int_{-\infty}^{+\infty} d\xi e^{-\frac{\left(\frac{x+\mu}{2} - \xi\right)^2}{(\sigma')^2}} \\
&= \sigma' \sqrt{\pi} e^{-\frac{(x-\mu)^2}{2\sigma^2}}
\end{aligned} \tag{S3}$$

where $\sigma = \sqrt{2}\sigma'$. It indicates that a "large" Gaussian function $\exp\left\{-\frac{(x-\mu)^2}{2\sigma^2}\right\}$ with a standard deviation of $\sigma$ can be represented by a convolution of two "small" Gaussian functions $\exp\left\{-\frac{(\xi-\mu)^2}{2(\sigma')^2}\right\}$ and $\exp\left\{-\frac{(x-\xi)^2}{2(\sigma')^2}\right\}$ with a standard deviation of $\sigma' = \sigma/\sqrt{2}$:

$$e^{-\frac{(x-\mu)^2}{2\sigma^2}} = \frac{1}{\sigma'\sqrt{\pi}} \int_{-\infty}^{+\infty} d\xi e^{-\frac{(\xi-\mu)^2}{2(\sigma')^2}} e^{-\frac{(x-\xi)^2}{2(\sigma')^2}} \tag{S4}$$

And a $D$-dimensional multivariate Gaussian function $G(\boldsymbol{x} - \boldsymbol{\mu})$ can be expressed as:

$$G(\boldsymbol{x} - \boldsymbol{\mu}) = e^{-\frac{1}{2}(\boldsymbol{x}-\boldsymbol{\mu})^T \boldsymbol{\Sigma}^{-1}(\boldsymbol{x}-\boldsymbol{\mu})}$$
$$= \frac{1}{C} \int_{-\infty}^{+\infty} d\boldsymbol{\xi} f(\boldsymbol{\xi} - \boldsymbol{\mu}) g(\boldsymbol{x} - \boldsymbol{\xi}) \quad (S5)$$
$$= \frac{1}{C} \int_{-\infty}^{+\infty} d\boldsymbol{\xi} e^{-(\boldsymbol{\xi}-\boldsymbol{\mu})^T \boldsymbol{\Sigma}^{-1}(\boldsymbol{\xi}-\boldsymbol{\mu})} e^{-(\boldsymbol{x}-\boldsymbol{\xi})^T \boldsymbol{\Sigma}^{-1}(\boldsymbol{x}-\boldsymbol{\xi})}$$

$$C = \sqrt{\pi^D \det(\boldsymbol{\Sigma})} \quad (S6)$$

where $C$ is the normalisation factor, $\boldsymbol{\Sigma}$ is the covariance matrix of the multivariate Gaussian function $\mathcal{G}(\boldsymbol{x} - \boldsymbol{\mu})$, $\boldsymbol{\Sigma}^{-1}$ is its inverse, and $\det(\boldsymbol{\Sigma})$ denotes the determinant of $\boldsymbol{\Sigma}$. But in most cases, we will only use the diagonal matrix as the covariance matrix, that is, use a $D$-dimensional vector as the standard deviation $\boldsymbol{\sigma} = \{\sigma_1, \sigma_2, \ldots, \sigma_D\}$:

$$G(\boldsymbol{x} - \boldsymbol{\mu}) = e^{-\frac{1}{2}\left\|\frac{\boldsymbol{x}-\boldsymbol{\mu}}{\boldsymbol{\sigma}}\right\|^2} = \frac{1}{C} \int_{-\infty}^{+\infty} d\boldsymbol{\xi} e^{-\frac{1}{2}\left\|\frac{\boldsymbol{\xi}-\boldsymbol{\mu}}{\boldsymbol{\sigma}'}\right\|^2} e^{-\frac{1}{2}\left\|\frac{\boldsymbol{x}-\boldsymbol{\xi}}{\boldsymbol{\sigma}'}\right\|^2} \quad (S7)$$

$$C = \sqrt{\pi^D \det(\mathrm{diag}(\boldsymbol{\sigma}))} = \prod_d^D (\sigma_d' \sqrt{\pi}) = \prod_d^D \left(\sigma_d \sqrt{\frac{\pi}{2}}\right) \quad (S8)$$

Therefore, we can fit the Gaussian-type repulsive potential $G[\boldsymbol{s}(\boldsymbol{R}); \boldsymbol{s}'(t)]$ of MetaD with a set of Gaussian-type basis-functions $g_i(\boldsymbol{s}) = \exp(-\frac{1}{2}\|\frac{\boldsymbol{s}-\boldsymbol{s}_i}{\boldsymbol{\sigma}'}\|^2)$ at the CV-grids $\{\boldsymbol{s}_i\}$ with spacing $\{\Delta S_i = \prod_d^D \Delta s_{i,d}\}$:

$$G[\boldsymbol{s}(\boldsymbol{R}); \boldsymbol{s}'(t)] = \omega(t) e^{-\frac{1}{2}\left\|\frac{\boldsymbol{s}(\boldsymbol{R})-\boldsymbol{s}'(t)}{\boldsymbol{\sigma}}\right\|^2} = \frac{\omega(t)}{C(t)} \sum_i \Delta S_i f_i(\boldsymbol{s}'(t)) g_i(\boldsymbol{s}(\boldsymbol{R})) \quad (S9)$$

$$f_i(\boldsymbol{s}'(t)) = e^{-\frac{1}{2}\left\|\frac{\boldsymbol{s}_i-\boldsymbol{s}'(t)}{\boldsymbol{\sigma}'}\right\|^2} = e^{-\left\|\frac{\boldsymbol{s}_i-\boldsymbol{s}'(t)}{\boldsymbol{\sigma}}\right\|^2} \quad (S10)$$

$f_i(\boldsymbol{s}'(t))$ is the weight coefficient of basis $g_i(\boldsymbol{s}(\boldsymbol{R}))$. We expect the integral of the fitted function to be equal to the integral of the original function, i.e.:

$$\int_{-\infty}^{+\infty} d\boldsymbol{s}\, e^{-\frac{1}{2}\left\|\frac{\boldsymbol{s}-\boldsymbol{s}'(t)}{\boldsymbol{\sigma}}\right\|^2} = \int_{-\infty}^{+\infty} d\boldsymbol{s}\, \frac{1}{C(t)} \sum_i \Delta S_i f_i(t)\, e^{-\frac{1}{2}\left\|\frac{\boldsymbol{s}-\boldsymbol{s}_i}{\boldsymbol{\sigma}'}\right\|^2}$$
$$\prod_d^D (\sigma_d \sqrt{2\pi}) = \frac{1}{C(t)} \prod_d^D (\sigma_d' \sqrt{\pi}) \sum_i \Delta S_i f_i(t) \quad (S11)$$

Thus, the normalisation factor $C(t)$ is:

$$C(t) = \frac{1}{(\sqrt{2})^D} \sum_i \Delta S_i f_i(t) \tag{S12}$$

If $s'(t)$ is not located at the margins of CV-grids $\{s_i\}$, then according to equation (S7), there is:

$$C(t) = \frac{1}{(\sqrt{2})^D} \sum_i \Delta S_i f_i(t) \approx C = \prod_d^D (\sigma_d' \sqrt{\pi}) \tag{S13}$$

Thus, the normalisation factor $C(t)$ can also be approximated as a constant $C$:

### B. Shift Factor

We obtain the sinking bias potential $V_{\text{sink}}(s(R);t)$ by subtracting a shift factor $v_{\text{shift}}(t)$ from each weighting parameter $k_i(t)$:

$$\begin{aligned}
V_{\text{sink}}(s(R);t) &= \sum_i^N (k_i(t) - v_{\text{shift}}(t)) \Delta S_i e^{-\frac{1}{2}\left\|\frac{s(R)-s_i}{\sigma'}\right\|^2} \\
&= \sum_i^N k_i(t) \Delta S_i e^{-\frac{1}{2}\left\|\frac{s(R)-s_i}{\sigma'}\right\|^2} - v_{\text{shift}}(t) \sum_i^N \Delta S_i e^{-\frac{1}{2}\left\|\frac{s(R)-s_i}{\sigma'}\right\|^2} \\
&= V(s(R),t) - v_{\text{shift}}(t) \Phi(s(R)) \\
&= V(s(R),t) + V_{\text{shift}}(s(R);t)
\end{aligned} \tag{S14}$$

$$\Phi(s) = \sum_i^N \Delta S_i e^{-\frac{1}{2}\left\|\frac{s(R)-s_i}{\sigma'}\right\|^2} \tag{S15}$$

where $\Phi(s)$ is the cumulative function, and $V_{\text{shift}}(s(R);t) = -v_{\text{shift}}(t)\Phi(s(R))$ is the shift potential. We expect the maximum value of the sinking bias potential $V_{\text{sink}}(s(R),t)$ to be a manually constant value $-E_{\text{depth}}$:

$$\begin{aligned}
\max\{V_{\text{sink}}(s;t)\} &= \max\{V(s;t) - v_{\text{shift}}(t)\Phi(s)\} \\
&= V_{\max}(t) - v_{\text{shift}}(t)\Phi[s(V_{\max}(t))] \\
&= -E_{\text{depth}}
\end{aligned} \tag{S16}$$

where $V_{\max}(t)$ is the maximum value of the original bias potential $V(s;t)$, and $s(V_{\max}(t))$ is the CV corresponding to $V_{\max}(t)$. Consider that the bias potential maximum $V_{\max}(t)$ is generally not located at the margins of the CV-grids $\{s_i\}$, so $\Phi[s(V_{\max}(t))]$ can be estimated as:

$$\Phi[s(V_{\max}(t))] = \sum_i^N \Delta S_i e^{-\frac{1}{2}\left\|\frac{s(V_{\max}(t))-s_i}{\sigma'}\right\|^2} \approx \prod_d^D (\sigma_d'\sqrt{2\pi}) = C' \tag{S17}$$

where $C' = \prod_d^D(\sqrt{2\pi}\sigma_d') = (\sqrt{2})^D C$ is a constant. Therefore, the shift factor $v_{\text{shift}}(t)$ is:

$$v_{\text{shift}}(t) = \frac{V_{\max}(t) + E_{\text{depth}}}{\Phi[\boldsymbol{s}(V_{\max}(t))]} \approx \frac{V_{\text{depth}}(t)}{\prod_d^D(\sigma'_d\sqrt{2\pi})} = \frac{V_{\text{depth}}(t)}{(\sqrt{2})^D C} \tag{S18}$$

where $V_{\text{depth}}(t) = V_{\max}(t) + E_{\text{depth}}$ is the depth to which the interior region of the bias potential $V(\boldsymbol{s}(\boldsymbol{R}),t)$.

## C. Boundary Effect

The bias force $\boldsymbol{F}_{\text{sink}}(\boldsymbol{R};t)$ of SinkMeta is:

$$\begin{aligned}\boldsymbol{F}_{\text{sink}}(\boldsymbol{R};t) &= -\frac{\partial V_{\text{sink}}(\boldsymbol{s}(\boldsymbol{R});t)}{\partial \boldsymbol{R}} \\ &= -\frac{\partial V(\boldsymbol{s}(\boldsymbol{R});t)}{\partial \boldsymbol{R}} - \frac{\partial V_{\text{shift}}(\boldsymbol{s}(\boldsymbol{R});t)}{\partial \boldsymbol{R}} \\ &= \boldsymbol{F}(\boldsymbol{R};t) + \boldsymbol{F}_{\text{shift}}(\boldsymbol{R};t)\end{aligned} \tag{S19}$$

where $F(\boldsymbol{R};t) = -\partial V(\boldsymbol{s}(\boldsymbol{R};t))/\partial \boldsymbol{R}$ is the original bias force before sinking, $\boldsymbol{F}_{\text{shift}}(\boldsymbol{R};t) = v_{\text{shift}}(t)\partial\big(\Phi(\boldsymbol{s}(\boldsymbol{R}))\big)/\partial \boldsymbol{R}$ is the shift force. Here, we use $\boldsymbol{S}$ to denote the space where the CV-grids $\{\boldsymbol{s}_i\}$ is located, then the cumulative function $\Phi(\boldsymbol{s}(\boldsymbol{R}))$ can be expresses as follow:

$$\Phi(\boldsymbol{s}) = \sum_i^N \Delta S_i e^{-\frac{1}{2}\left\|\frac{\boldsymbol{s}-\boldsymbol{s}_i}{\sigma'}\right\|^2} \approx \int_{\boldsymbol{S}} d\boldsymbol{\xi}\, e^{-\frac{1}{2}\left\|\frac{\boldsymbol{\xi}-\boldsymbol{s}}{\sigma'}\right\|^2} \tag{S20}$$

Let us first consider the case of one-dimensional CV $s$. In this situation, $\Phi(s)$ can be approximated as the integral from the lower bound $s_{\min}$ to the upper bound $s_{\max}$ of the CV grids $\{s_i\}$:

$$\begin{aligned}\Phi(s) &\approx \int_{s_{\min}}^{s_{\max}} d\xi\, e^{-\frac{1}{2}\left(\frac{\xi-s}{\sigma'}\right)^2} \\ &= \frac{\sigma'\sqrt{2\pi}}{2}\left[\text{erf}\left(\frac{s_{\max}-s}{\sigma'\sqrt{2}}\right) - \text{erf}\left(\frac{s_{\min}-s}{\sigma'\sqrt{2}}\right)\right]\end{aligned} \tag{S21}$$

Then, the shift potential $V_{\text{shift}}(s;t)$ is:

$$\begin{aligned}V_{\text{shift}}(s;t) &= -v_{\text{shift}}(t)\Phi(s) \\ &\approx \frac{1}{2}V_{\text{depth}}(t)\left[\text{erf}\left(\frac{s-s_{\max}}{\sigma'\sqrt{2}}\right) - \text{erf}\left(\frac{s-s_{\min}}{\sigma'\sqrt{2}}\right)\right]\end{aligned} \tag{S22}$$

As a result, the shift potential $V_{\text{shift}}(s,t)$ is equivalent to forming "cliffs" at the margins of the CV grids $\{s_i\}$ with the lower and upper restraining potentials $V_{\text{rest}}^{\text{lower}}(s;t)$ and $V_{\text{rest}}^{\text{upper}}(s;t)$:

$$V_{\text{rest}}^{\text{lower}}(s;t) = -\frac{1}{2}V_{\text{depth}}(t)\left[1 + \text{erf}\left(\frac{s - s_{\min}}{\sigma'\sqrt{2}}\right)\right] \tag{S23}$$

$$V_{\text{rest}}^{\text{upper}}(s;t) = -\frac{1}{2}V_{\text{depth}}(t)\left[1 - \text{erf}\left(\frac{s - s_{\max}}{\sigma'\sqrt{2}}\right)\right] \tag{S24}$$

And the lower and upper restraining forces $F_{\text{rest}}^{\text{lower}}(s;t)$ and $F_{\text{rest}}^{\text{upper}}(s;t)$ are:

$$F_{\text{rest}}^{\text{lower}}(s;t) = -\frac{\partial\left(V_{\text{rest}}^{\text{lower}}(s;t)\right)}{\partial s} = \frac{V_{\text{depth}}(t)}{\sigma'\sqrt{2\pi}}e^{-\frac{1}{2}\left(\frac{s-s_{\min}}{\sigma'}\right)^2} \tag{S25}$$

$$F_{\text{restr}}^{\text{upper}}(s;t) = -\frac{\partial\left(V_{\text{rest}}^{\text{upper}}(s;t)\right)}{\partial s} = -\frac{V_{\text{depth}}(t)}{\sigma'\sqrt{2\pi}}e^{-\frac{1}{2}\left(\frac{s-s_{\max}}{\sigma'}\right)^2} \tag{S26}$$

The upper and lower restraining potentials $V_{\text{rest}}^{\text{lower}}(s;t)$ and $V_{\text{rest}}^{\text{upper}}(s;t)$ will confine the CV $s$ to the interior of the grids $\{s_i\}$.

And the shift bias $V_{\text{shift}}^{\text{inter}}(s;t)$ at the grid interior equal to a constant value $-V_{\text{depth}}(t)$, so the bias force $\boldsymbol{F}_{\text{sink}}^{\text{inter}}(\boldsymbol{R};t)$ of SinkMeta at this region is approximated to the original bias force $\boldsymbol{F}(\boldsymbol{R};t)$ before sinking:

$$\boldsymbol{F}_{\text{sink}}^{\text{inter}}(\boldsymbol{R};t) = \boldsymbol{F}(\boldsymbol{R};t) - \frac{\partial V_{\text{shift}}^{\text{inter}}(s;t)}{\partial \boldsymbol{R}} \approx \boldsymbol{F}(\boldsymbol{R};t) \tag{S27}$$

The case of multidimensional CV $\boldsymbol{s}(\boldsymbol{R})$ is analogous to that of one-dimensional. The restraining potential $V_{\text{shift}}^{\text{margin}}(\boldsymbol{s}(\boldsymbol{R}))$ at the margins of the CV-grids $\{\boldsymbol{s}_i\}$ will prevent CVs $\boldsymbol{s}(\boldsymbol{R})$ from escaping to the exterior of the grids $\{\boldsymbol{s}_i\}$, while CVs trapped in the grid interior will be under the same bias force $\boldsymbol{F}(\boldsymbol{R};t)$ as in the regular WT-MetaD approach.